\documentclass[reprint,amsmath,showkeys,amssymb,superscriptaddress,eqsecnum]{revtex4-2}
\usepackage{CJKutf8}
\usepackage{mathrsfs}
\usepackage{bbm}
\usepackage{graphicx}
\usepackage{dcolumn}
\usepackage{bm}
\usepackage{float}
\usepackage{color}
\usepackage{multirow}
\usepackage{enumerate}
\usepackage{BOONDOX-cal}
\begin{document}
\begin{CJK*}{UTF8}{}
\CJKfamily{gbsn}
\title{Nucleon-pair approximation with matrix representation}
\author{Y. Lei ({\CJKfamily{gbsn}雷杨})}
\affiliation{School of National Defense Science and Technology, Southwest University of Science and Technology, Mianyang 621010, China}
\author{Y. Lu ({\CJKfamily{gbsn}路毅})}
\email[corresponding author: ]{luyi@qfnu.edu.cn}
\affiliation{College of Physics and Engineering, Qufu Normal University, 57 Jingxuan West Road, Qufu, Shandong 273165, China}

\author{Y. M. Zhao ({\CJKfamily{gbsn}赵玉民})}
\affiliation{Shanghai Key Laboratory of Particle Physics and Cosmology, School of Physics and Astronomy, Shanghai Jiao Tong University, Shanghai 200240, China}
\affiliation{Collaborative Innovation Center of IFSA (CICIFSA), Shanghai Jiao Tong University, Shanghai 200240, China}
\date{\today}

\begin{abstract}
In this paper, we propose an approach of the nucleon-pair approximation (NPA), in which the collective nucleon pairs are represented in terms of antisymmetric matrices, and commutations between nucleon pairs are given by using matrix multiplication that avoids angular-momentum couplings and recouplings. Therefore the present approach significantly simplifies the NPA computation. Furthermore, it is formulated on the same footing with and without isospin.
\end{abstract}
\maketitle
\section{Introduction}\label{sec-int}

Pairing phenomenon is observed in a wide range of quantum many-body systems and scales, from finite nuclei to neutron star \cite{pair-rev-1}. 
Pairing correlation in the nucleus is originated from the short-range and attractive nuclear interaction between like nucleons; in this case, two nucleons have large spatial overlaps and achieve low energy. Such two nucleons are called one nucleon pair. This correlation plays an important role in low-lying states, in particular, of semi-magic nuclei.

The theoretical formulation of pairing correlation can be traced back to the seniority scheme in a single-$j$ shell \cite{sen-1,sen-2,sen-3}, suggested by Racah about eighty years ago. In 1957, Bardeen, Cooper, and Schrieffer proposed the BCS theory of superconductivity in metal conductors at low temperatures \cite{bcs-1,bcs-2}. This approach was introduced to nuclear physics to describe the pairing correlation in Refs. \cite{bcs2nucl-bohr,bcs2nucl-2,bcs2nucl-3}. About fifty years ago, Talmi generalized the seniority scheme for semi-magic nuclei with many-$j$ shells \cite{gen-sen-1,gen-sen-2}. Richardson provided an exact numerical solution to the simplified pairing model, which was applied to various strongly correlated many-body systems due to its intimate link to several solvable pair models \cite{pair-rev-2}. 
As an extension of the BCS and generalized seniority scheme, the broke-pair approximation \cite{bpa-rev} adopts collective-$S$ (spin-zero) nucleon pairs and a very few broken (spin non-zero) nucleon pairs, in studies of the low-energy states of semi-magic nuclei.
In the nineteen seventies, Arima and Iachello proposed the interacting boson model (IBM) \cite{ibm} in which the building blocks, the $sd$ bosons are mappings of $SD$ (spin zero and two) nucleon pairs \cite{ibm-sm-fou}. Enlightened by the IBM, Ginocchio proposed a fermionic model in which $SD$ nucleon pairs follow dynamical symmetries \cite{ginocchio-1, ginocchio-2}, and this Ginocchio model was further developed by Wu and collaborators, called the fermion dynamical symmetry model (FDSM) \cite{fdsm}.

In nineteen nineties, Chen proposed the Wick theorem for coupled fermion pairs \cite{npa-for-chen-1,npa-for-chen-2}, and based on this novel technique Chen established the nucleon-pair shell model (NPSM) \cite{npa-for-chen-3}, also called the nucleon-pair approximation (NPA) in literature \cite{npa-rev-zhao}. This approach was refined to treat both even and odd nuclei in a more sagacious way \cite{npa-for-zhao}. In recent years the isospin symmetry and particle-hole configuration were considered in the NPA \cite{npa-iso-fu,npa-ph-cheng}. However, computations become heavy if the valence proton number and/or valence neutron pair number is larger than eight. Therefore studies of rotational motion for heavy nuclei are prohibitively challenging in these NPA approaches.

In the last two decades, the $m$-scheme shell model calculations become more and more affordable than the $j$-scheme shell-model calculations, with the rapid development of computer memory. In Ref. \cite{npa-cal-yos-1} Higashiyama and Yoshinaga expanded the NPA basis states in terms of $m$-scheme basis states, and performed the NPA calculation by using $m$-scheme shell model code. Very recently, He and his collaborators performed the NPA calculations in terms of collective nucleon pairs but without angular-momentum coupling between pairs in Ref. \cite{mnpa-luo}, and called their approach nucleon-pair shell model in $m$ scheme. As the approach of Ref. \cite{mnpa-luo} does not resort to the angular-momentum couplings of nucleon pairs, and thus is more realizable than the transitional NPA calculations of Refs. \cite{npa-for-chen-3,npa-rev-zhao,npa-cal-yos-1}; Yet, the commutators between nucleon pair operators and one-body (or two-body) operators remain to be the same as the NPA approach of Refs. \cite{npa-for-chen-3,npa-rev-zhao}, and thus still suffer from sophisticated angular-momentum couplings and re-couplings.

It is therefore the purpose of this paper to propose an approach of the NPA, in which one resorts neither to angular-momentum couplings and re-couplings in the evaluation of commutators between collective nucleon pairs, nor to the couplings and re-couplings of ``new" nucleon pairs in the basis states. Namely, angular-momentum couplings and re-couplings are avoided from the beginning to the end of the computation. 

Towards this goal, we adopt the matrix representation of collective nucleon pairs. Matrix representation has been developed many years ago by Silvestre-Brac and Piepenbring in studies of multi-phonon states \cite{mpm-1,mpm-3,mpm-odd}; Ginocchio and Johnson derived a similar formalism based on generating functions \cite{ioa-mapping-ginocchio-johnson}; Mizusaki and Otsuka adopted a matrix representation in studies of the shell-model foundation of the IBM \cite{ibm-pair-o6}. Otsuka $et.~al.$ also used it to formulate pair truncation of Monte Carlo shell model \cite{mcsm-pair}.

In this paper, we develop this matrix representation for collective nucleon pairs and derive commutators, overlaps of the NPA basis states, and matrix elements of the shell-model Hamiltonian. The NPA with and without isospin symmetry, for even-even nuclei and odd-mass nuclei, is presented in the unified formulation. By using this version of the NPA, numerical calculations of rotational states for heavy nuclei are now readily realizable.

This paper is organized as follows. In Sec. \ref{sec-for} we define the matrix representation of collective nucleon pairs, and one-body and two-body operators, and present commutators between nucleon pairs and/or operators. In Sec. \ref{sec-for-2} we derive overlaps and the matrix elements of the shell model Hamiltonian, for both even and odd systems. In Sec. \ref{sec-mat-npa} we discuss matrix representation of ``traditional" NPA \cite{npa-for-chen-3,npa-rev-zhao}.
In Sec. \ref{sec-alg} we demonstrate the computation power of our new approach in this paper. In Sec. \ref{sec-sum} we summarize this paper.

\section{matrix-represented commutations}\label{sec-for}

In this section, we describe matrix representation of collective nucleon pairs, one-body and two-body operators, and unpaired particles, as well as commutators between collective nucleon pairs and operators.

\subsection{pair and one-body operator}\label{subsec-bas-def}
We define collective nucleon-pair creation and annihilation operators as below,
\begin{equation}\label{eq-pair-def}
\begin{aligned}
\hat{P}^{\dagger}=&\frac{1}{2}\sum_{\alpha \beta} p(\alpha \beta) \hat{c}^{\dagger}_\alpha \hat{c}^{\dagger}_\beta,\\
\hat P=& (\hat P^{\dagger})^{\dagger}=-\frac{1}{2}\sum_{\alpha \beta} p(\alpha \beta) \hat{c}_\alpha \hat{c}_\beta,
\end{aligned}
\end{equation}
where $\alpha$ and $\beta$ denote orthonormal single-particle basis, and $\hat{c}^{\dagger} ~(\hat{c})$ is a fermion creation (annihilation) operator. For example, in spherical basis, the $\alpha / \beta$ corresponds to the $\{nljm\}$ quantum numbers or $\{nljm\tau\}$ with an extra isospin projection, $\tau$. $p(\alpha \beta)$ are pair structural coefficients, with anti-symmetry $ p(\alpha \beta) = - p(\beta \alpha)$; these $p(\alpha \beta)$ are treated to be matrix elements of matrix $p$.

Similarly, a general one-body operator is defined as
\begin{equation}\label{eq-q}
\hat{Q} = \sum_{\alpha \beta} q(\alpha \beta) \hat{c}^{\dagger}_\alpha \hat{c}_{\beta},
\end{equation}
where coefficients $q(\alpha \beta)$ also constructs a matrix $q$. Accordingly, the conjugate operator of $\hat Q$, i.e., $\hat Q^{\dagger}$, has its structural matrix as $q^\top$, i.e., the transpose matrix of $q$.

In odd-nucleon system, an unpaired nucleon is represented by using a linear combination of single-particle basis state,
\begin{equation}
\hat{a}^\dagger | 0 \rangle = \sum_\alpha a_\alpha \hat{c}^\dagger_\alpha | 0 \rangle,
\label{eqn:single-particle-operator}
\end{equation}
where $a_\alpha$ are structural coefficients of $\hat{a}^\dagger$; in case that the unpaired nucleon occupies given single-particle level $\alpha$, $a_\alpha =1$ and $a_{\alpha'} =0$ $(\alpha' \neq \alpha)$. Similarly, the coefficients $a_\alpha$ construct a column vector $\vec{a}$.

Our nucleon-pair basis states of matrix representation is given by
\begin{equation}\label{eqn:basis}
\left.\left.\hat{P}^{\dagger}_0 \hat{P}^{\dagger}_1 \hat{P}^{\dagger}_2 \cdots \hat{P}^{\dagger}_N \right|0\right\rangle
\end{equation}
with
\begin{equation*}
\hat P^{\dagger}_0=\left\{
\begin{array}{cc}
1&{\rm ~in~even~system}\\
\hat a^{\dagger} &{\rm ~in~odd~system}
\end{array}
\right..
\end{equation*}

\subsection{Commutations between pair and one-body operators}\label{sec-com}

In this subsection, we present key commutators between one nucleon-pair operator and one-body operator in the decoupled form, as defined in the last subsection.

The commutator between nucleon-pair annihilation operator and nucleon-pair creation operator is easily
obtained as follows.
\begin{equation}
\left[ \hat{P}_1, \hat{P}^{\dagger}_2 \right] = - \frac{1}{2} {\rm tr}\left( p_2 p_1 \right) + \hat{\mathcal Q}.
\label{eqn:[A,A+]}
\end{equation}
where $p_1, p_2$ are structure matrices of $\hat{P}_1, \hat{P}_2$, ${\rm tr}( p_2 p_1 )$ means the trace of matrix product $ p_2 p_1 $, and $\hat{\mathcal Q}$ is a one-body operator 
\begin{eqnarray}
\hat{\mathcal Q} = \sum_{\alpha \beta} {\mathcal q}(\alpha \beta) \hat{c}^{\dagger}_\alpha \hat{c}_{\beta}~, \nonumber 
\end{eqnarray}
with structural matrix 
\begin{eqnarray}
\mathcal q = p_2 p_1~. \nonumber 
\end{eqnarray}

Similarly, the commutator between nucleon-pair annihilation operator, $\hat{P}_3$, and one-body operator, $\hat{Q}$, yields another nucleon-pair annihilation operator $\hat P_4$,
\begin{equation}\label{eqn:[A,Q]-matrix-rep}
\hat{P}_4 = [\hat{P}_3, \hat{Q} ]~, ~~ {\rm ~with~} p_4 = p_3 q + q^\top p_3~,
\end{equation}
where $p_3$ and $p_4$ is structural coefficient matrices of $\hat{P}_3$ and $\hat P_4$, respectively.

The single-particle operators follow the relation as below,
\begin{equation}
\hat{a} \hat{b}^\dagger
= \left\{ \hat{a}, \hat{b}^\dagger \right\}
- \hat{b}^{\dagger} \hat{a} = \vec{a} \cdot \vec{b} - \hat{\mathbb{Q}},
\label{eqn:single-particle-operator-commutator}
\end{equation}
where $\vec{a} \cdot \vec{b}$ is the inner product of vectors $\vec{a}$ and $\vec{b}$, and $\hat{\mathbb{ Q}}$ is a one-body operator with structural matrix $\mathbb{q}=\vec{b} \vec{a}^\top$.

The commutator between single-particle annihilation operator, $\hat{a}$, and one-body operator, $\hat{Q}$, yields another single-particle annihilation operator, $\hat{d}$,
\begin{equation}
\hat{d} = [ \hat{a}, \hat{Q} ] = \sum_{\alpha} d_{\alpha} \hat{c}_{\alpha} ~, ~~ {\rm ~with~} \vec d =q^\top \vec{a}~,
\label{eqn:[a,Q]}
\end{equation}
where vector $\vec d$ represents structural coefficients $\hat{d}$ operator. 

The commutator between single-particle annihilation operator, $\hat{a}$, and pair creation operator, $\hat P^{\dagger}$, produces a single-particle creation operator, $\hat{d}^{\dagger}$,
\begin{equation}
\hat{d}^{\dagger} = [ \hat{a}, \hat P^{\dagger}] = \sum_{\alpha} d_{\alpha} \hat{c}^{\dagger}_{\alpha} ~, ~~{\rm ~with~} \vec d =-2p \vec a~,
\label{eqn:[a,A]}
\end{equation}
where, again, $\vec d$ represents structural coefficients $\hat{d}^{\dagger}$ operator.

\begin{widetext}

\subsection{Commutators of $N$ nucleon pair operators}\label{sec-com}

By using Eqs. (\ref{eqn:[A,A+]}) to (\ref{eqn:[a,A]}), one easily derives commutator between $N$ nucleon-pair annihilation operator and one-body operator,
\begin{equation}\label{eq-con-ppq}
\left[\hat P_0\hat P_1\right.\left.\cdots \hat P_N,\hat Q\right]=\sum_{k=0}^N \hat P_0\cdots \hat P_{k-1}\left[P_k,\hat Q\right]\hat P_{k+1}\cdots \hat P_N,
\end{equation}
where
\begin{equation}\label{eq-ppq-ope-def}
\left[\hat P_k,\hat Q\right]=\left\{
\begin{array}{llll}
\hat{\mathcal P}_k&{\rm with}&\mathcal p_k=p_kq+q^{\top}p_k&k\neq 0 ~, \\
\hat d&{\rm with}&\hat P_0=\hat a {\rm ~and~} \vec d=q^\top \vec a&k=0 ~. 
\end{array}
\right.
\end{equation}
Similarly, one derives commutators between $N$ nucleon-pair annihilation operator and one nucleon-pair creation operator,
\begin{eqnarray}\label{eq-ppp-pre}
\left[\hat P_0\hat P_1\cdots \hat P_N,\hat{\mathbb{ P}}^{\dagger} \right]&=&\sum_{k=0}^N \hat P_0\cdots \hat P_{k-1}\left[P_k,\hat{ \mathbbm{P}}^{\dagger}\right]\hat P_{k+1}\cdots \hat P_N\nonumber\\
&=&\sum_{k=1}^N -\frac{1}{2}{\rm tr}(\mathbbm{ p} p_k)\hat P_0\cdots \hat P_{k-1}\hat P_{k+1}\cdots \hat P_N\nonumber\\
&&+\sum_{k=1}^N \hat P_0\cdots \hat P_{k-1}\hat{\mathcal Q}_k\hat P_{k+1}\cdots \hat P_N+\left.\hat h^{\dagger}\hat P_1\cdots \hat P_N\right|_{\rm in~odd~system},
\end{eqnarray}
where the one-body operator $\hat{\mathcal{Q}}_k$ has structural matrix as $\mathcal q_k= \mathbbm{ p}p_k$, and the single-particle creation operator $\hat h^{\dagger}$ has structural coefficients represented by $\vec h=-2 \mathbbm{p} a$. The $\hat P_0\cdots \hat P_{k-1}\hat{\mathcal Q}_k$ term in the second term of right-hand side of the above equation is re-organized as below,
\begin{equation}\label{eq-ppk-1q}
\hat P_0\cdots \hat P_{k-1}\hat{\mathcal Q}_k=\left[\hat P_0\cdots \hat P_{k-1},\hat{\mathcal Q}_k\right]+\hat{\mathcal Q}_k\hat P_0\cdots \hat P_{k-1}
=\sum_{i=0}^{k-1}\hat P_0\cdots \hat P_{i-1}\left[\hat P_i,\hat{\mathcal Q}_k\right]P_{i+1}\cdots \hat P_{k-1}+\hat{\mathcal Q}_k\hat P_0\cdots \hat P_{k-1},
\end{equation}
where
\begin{equation}\label{eq-ppp-ope-def}
\left[\hat P_i,\hat{\mathcal Q}_k\right]=\left\{
\begin{array}{llll}
\hat{\mathcal P}_{i,k}&{\rm with}&\mathcal{ p}_{i,k}=p_i \mathbbm{p} p_k+p_k \mathbbm{p} p_i&i\neq 0 ~,\\
\hat e_k&{\rm with}&\hat P_0=\hat a{\rm ~and ~}\vec e_k=p_k \mathbbm{p} \vec a&i=0 ~.
\end{array}
\right.
\end{equation}
From Eqs. (\ref{eq-ppp-pre}) and (\ref{eq-ppk-1q}), one has
\begin{eqnarray}\label{eq-con-ppp}
\left[\hat P_0\hat P_1\cdots \hat P_N,\hat{ \mathbbm{P}}^{\dagger} \right]&=&\sum_{k=1}^N -\frac{1}{2}{\rm tr}(\mathbbm{p} p_k)\hat P_0\cdots \hat P_{k-1}\hat P_{k+1}\cdots \hat P_N+\sum_{k=2}^N\sum_{i=1}^{k-1}\hat P_0\cdots \hat P_{i-1}\hat{\mathcal P_{i,k}}\hat P_{i+1}\cdots\hat P_{k-1}\hat P_{k+1}\cdots \hat P_N\nonumber\\
&&+\left.\sum_{k=1}^N \hat e_k\cdots \hat P_{k-1}\hat P_{k+1}\cdots \hat P_N\right|_{\rm odd~system}\nonumber\\
&&+\sum_{k=1}^N\hat{\mathcal Q}_k \hat P_0\cdots \hat P_{k-1}\hat P_{k+1}\cdots \hat P_N+\left.\hat d^{\dagger}\hat P_1\cdots \hat P_N\right|_{\rm odd~system},
\end{eqnarray}
where $\mathcal p_{i,k}$ and $\vec e_k$ are given in Eq. (\ref{eq-ppp-ope-def}).

The two-body interaction operator are written in the form of $\hat A^\dagger \hat B$, where $\hat A$ is collective nucleon-pair creation operator with structural matrix $p_A$, and $B$ is collective nucleon-pair annihilation operator. From Eq. (\ref{eq-con-ppp}), one has
\begin{eqnarray}\label{eq-con-pppp}
\left[\hat P_0\hat P_1\cdots \hat P_N,\hat{A}^{\dagger}\hat B \right]&=&\sum_{k=1}^N -\frac{1}{2}{\rm tr}(p_A p_k)\hat P_0\cdots \hat P_{k-1}\hat B\hat P_{k+1}\cdots \hat P_N\nonumber\\
&&+\sum_{k=2}^N\sum_{i=1}^{k-1}\hat P_0\cdots \hat P_{i-1}\hat{\mathcal P_{i,k}}\hat P_{i+1}\cdots\hat P_{k-1}\hat B\hat P_{k+1}\cdots \hat P_N\nonumber\\
&&+\left.\sum_{k=1}^N \hat e_k\cdots \hat P_{k-1}\hat B\hat P_{k+1}\cdots \hat P_N\right|_{\rm in~odd~system}\nonumber\\
&&+\sum_{k=1}^N\hat{\mathcal Q}_k \hat P_0\cdots \hat P_{k-1}\hat B\hat P_{k+1}\cdots \hat P_N+\left.\hat d^{\dagger}\hat P_1\cdots \hat P_N\right|_{\rm odd~system},
\end{eqnarray}
where the structural coefficients of operators $\hat{\mathcal P_{i,k}}$, $\hat e_k$ and $\hat{\mathcal Q}_k$ are the same as in Eq. (\ref{eq-ppp-ope-def}), with replacement of $\mathbbm{p}$ by using $p_A$.

The two-body operator in particle-hole channel is written in the form of $\hat Q \hat Q^{\dagger}$. One readily has
\begin{eqnarray} \label{N-pair-QQ-0}
\left[\hat P_0\hat P_1\cdots \hat P_N,\hat Q \hat Q^{\dagger}\right]&=&\left[\hat P_0\hat P_1\cdots \hat P_N,\hat Q \right]\hat Q^{\dagger}+\hat Q \left[\hat P_0\hat P_1\cdots \hat P_N,\hat Q^{\dagger} \right]\nonumber\\
&=&\left[\left[\hat P_0\hat P_1\cdots \hat P_N,\hat Q \right],\hat Q^{\dagger}\right]+\hat Q^{\dagger}\left[\hat P_0\hat P_1\cdots \hat P_N,\hat Q \right]+\hat Q \left[\hat P_0\hat P_1\cdots \hat P_N,\hat Q^{\dagger} \right].
\end{eqnarray}
Iterative application of Eq. (\ref{eq-con-ppq}) to the above commutator yields
\begin{eqnarray}
\left[\left[\hat P_0\hat P_1\cdots \hat P_N,\hat Q \right],\hat Q^{\dagger}\right]&=&\left[\sum_{k=0}^{N}\hat P_0\cdots \hat P_{k-1}\left[\hat P_{k+1},\hat Q\right]\hat P_{k+1}\cdots \hat P_N,\hat Q^{\dagger}\right]\nonumber\\
&=&\sum_{k,i=0}^{N,k\neq i}\hat P_0 \cdots
\left\{
\begin{array}{cc}
\hat{P}_{i-1} \left[\hat P_i,\hat Q^{\dagger }\right] \hat{P}_{i+1} \cdots \hat{P}_{k-1} \left[\hat P_k,\hat Q\right] \hat{P}_{k+1}&i<k\\
\hat{P}_{k-1} \left[\hat P_k,\hat Q\right] \hat{P}_{k+1}\cdots \hat{P}_{i-1}\left[\hat P_i,\hat Q^{\dagger }\right] \hat{P}_{i+1} &k<i\\
\end{array}
\right\}
\cdots \hat P_N\nonumber\\
&&+\sum_{k=i=0}^N \hat P_0 \cdots \hat{P}_{k-1} \left[\left[\hat P_k,\hat Q\right],\hat Q^{\dagger}\right] \hat{P}_{k+1}\cdots \hat P_N,
\end{eqnarray}
where
\begin{equation}\label{eq-ppqq-ope-def}
\begin{aligned}
&\left[\hat P_k,\hat Q \right]=\left\{
\begin{array}{llll}
\hat{\mathcal P}_k&{\rm with}&\mathcal p_k=p_kq+q^\top p_k&k\neq 0\\
\hat e&{\rm with}&\hat P_0=\hat a {\rm ~and~}\vec e=q^\top \vec a&k= 0\\
\end{array}
\right.,~
\left[\hat P_i,\hat Q^{\dagger} \right]=\left\{
\begin{array}{llll}
\hat{\mathcal{P}}_i&{\rm with}& \mathcal{p}_i=p_iq^\top+q p_i&i\neq 0\\
\hat f&{\rm with}&\hat P_0=\hat a {\rm ~and~}\vec f=q^\top \vec a&i= 0\\
\end{array}
\right.,\\
&\left[\left[\hat P_k,\hat Q \right],\hat Q^{\dagger}\right]=\left\{
\begin{array}{llll}
\hat{\mathfrak P}_k&{\rm with}&\mathfrak p_k=p_k q q^\top + q^\top p_k q^\top + q p_k q + q q^\top p_k&k\neq 0\\
\hat g&{\rm with}&\hat P_0=\hat a {\rm ~and~}\vec g=qq^\top \vec a&k= 0\\
\end{array}
\right..
\end{aligned}
\end{equation}
Substituting this result into Eq. (\ref{N-pair-QQ-0}), one has
\begin{eqnarray}\label{eq-con-ppqq}
\left[\hat P_0\hat P_1\cdots \hat P_N,\hat Q \hat Q^{\dagger}\right]&=&
\sum_{k,i=1}^{N,k\neq i}\hat P_0 \cdots
\left\{
\begin{array}{cc}
\hat{P}_{i-1} \hat{\mathcal P}_i \hat{P}_{i+1} \cdots \hat{P}_{k-1} \hat{\mathcal P}_k \hat{P}_{k+1}&i<k\\
\hat{P}_{k-1} \hat{\mathcal P}_k \hat{P}_{k+1}\cdots \hat{P}_{i-1}\hat{\mathcal P}_i \hat{P}_{i+1} &k<i\\
\end{array}
\right\}
\cdots \hat P_N\nonumber\\
&&+\sum_{k=i=1}^N \hat P_0\cdots\hat{P}_{k-1} \hat{\mathfrak P}_k \hat{P}_{k+1}\cdots \hat P_N\nonumber\\
&&\left.+\left\{\sum_{i=1}^{N}\hat e \hat{P}_{k-1} \hat{\mathcal P}_k \hat{P}_{k+1}\cdots \hat P_N+\sum_{k=1}^{N}\hat f \hat{P}_{i-1} \hat{\mathcal P}_i \hat{P}_{i+1}\cdots \hat P_N+\hat g\cdots \hat P_N\right\}\right|_{\rm odd~ system}\nonumber\\
&&+\hat Q^{\dagger}\left[\hat P_0\hat P_1\cdots \hat P_N,\hat Q \right]+\hat Q \left[\hat P_0\hat P_1\cdots \hat P_N,\hat Q^{\dagger} \right].
\end{eqnarray}

\section{overlaps of basis states and matrix elements of Hamiltonian}\label{sec-for-2}

In this section, we will present general formulas for the overlaps of basis states and the matrix elements of the Hamiltonian matrix elements. Some formulas are in similar form as those in Refs. \cite{mpm-3,mpm-odd}; the essential difference is that Refs. \cite{mpm-3,mpm-odd} treated phonon states with quasi-particle pairs, while here we adopt the $m$-scheme NPA basis states constructed from valence-nucleon pairs. As in the previous NPA formulation \cite{npa-for-chen-3,npa-for-zhao,npa-iso-fu,npa-ph-cheng}, in this section we first discuss the overlaps of basis states for systems without unpaired nucleon. The overlaps for an odd-mass nucleus is then presented in terms of matrix elements of one-body operators in the basis states of its even-even neighboring nucleus, and matrix elements of the shell-model Hamiltonian for an odd-mass nucleus are written in terms of overlaps of basis states for this odd-mass nucleus. Therefore, the evaluation of overlaps for basis states without unpaired nucleon is the key computation, as in the previous versions of the NPA in Refs. \cite{npa-for-chen-3,npa-for-zhao,npa-iso-fu,npa-ph-cheng,mnpa-luo}.

\subsection{Even-nucleon system}\label{sec-ove-even}

Let us first come to the the overlap $\left\langle 0\left|\hat{P}_1 \cdots \hat{P}_N \left| \hat{\mathbb P}^\dagger_1 \cdots \hat{\mathbb P}^\dagger_N\right.\right |0\right\rangle$. By using Eq. (\ref{eq-con-ppp}), one has
\begin{eqnarray}\label{eq-ove-re}
\left\langle 0\left|\hat{P}_1 \cdots \hat{P}_N \left| \hat{\mathbb P}^\dagger_1 \cdots \hat{\mathbb P}^\dagger_N\right.\right| 0\right\rangle &=&- \sum^N_{k=1} \frac{1}{2} {\rm tr}( \mathbbm{p}_N p_k )\left\langle 0\left|\hat{P}_1 \cdots \hat{P}_{k-1} \hat{P}_{k+1} \cdots \hat{P}_{N} \left| \hat{\mathbb P}^\dagger_1 \cdots \hat{\mathbb P}^\dagger_{N-1} \right.\right|0\right\rangle\nonumber\\
&& + \sum^N_{k=1}\sum^{k-1}_{i=1} \left\langle 0\left|\hat{P}_1 \cdots\hat{P}_{i-1} \hat{ \mathcal P }_{i,k} \hat{P}_{i+1} \cdots \hat{P}_{k-1} \hat{P}_{k+1}\cdots \hat{P}_{N} \left| \hat{ \mathbb P}^\dagger_1 \cdots \hat{ \mathbb P }^\dagger_{N-1} \right.\right|0\right\rangle,
\end{eqnarray}
where $\hat{\mathcal P}_{i,k}$ is given in Eq. (\ref{eq-ppp-ope-def}). The above formula of $N$-pair overlap can be used recursively, and thus is reduced to a sum of overlaps for $N=1$,
\begin{eqnarray}
\left\langle 0\left|\hat P\left|\hat {\mathbb P}^\dagger\right.\right|0\right\rangle=\left\langle0\left|\left[\hat P,\hat {\mathbb P}^\dagger\right]\right|0\right \rangle = - \frac{1}{2}{\rm tr}(\mathbbm{p} p)~, \nonumber
\end{eqnarray}
where $p$ and $\mathbbm{p}$ are structural matrices of $\hat P$ and $\hat {\mathbb P}^\dagger$, respectively.

By using Eq. (\ref{eq-con-ppq}), one obtains the matrix element of one-body operator, $\hat Q$, in the nucleon-pair basis,
\begin{eqnarray}\label{eqn:ME-Q}
\left\langle 0\left| \hat{P}_1 \cdots \hat{P}_N \left| \hat{Q} \right| \hat{\mathbbm{P}}^\dagger_1 \cdots \hat{\mathbbm{P}}^{\dagger}_N \right|0\right\rangle&=&\left\langle 0\left| \left[\hat{P}_1 \cdots \hat{P}_N ,\hat{Q}\right] \hat{\mathbbm{P}}^\dagger_1 \cdots \hat{\mathbbm{P}}^{\dagger}_N \right|0\right\rangle\nonumber\\
&=&\sum_{k=1}^{N}\left \langle 0\left| \hat{P}_1 \cdots \hat{P}_{k-1} \hat{\mathcal{P}}_k \hat{P}_{k+1} \cdots \hat{P}_N\left |
\hat{ \mathbbm{P}}^{\dagger}_1 \cdots \hat{\mathbbm{P}}^{\dagger}_N \right.\right|0\right\rangle,
\end{eqnarray}
where $\hat{\mathcal P}_k$ is given in Eq. (\ref{eq-ppq-ope-def}).

From Eq. (\ref{eq-con-pppp}), one obtains the matrix element of two-body interaction operator in the particle-particle channel, $\hat{A}^\dagger \hat{B}$,
\begin{eqnarray}\label{eqn:ME-AA}
\left\langle 0\left|\hat{P}_1 \cdots \hat{P}_N \left| \hat{A}^\dagger \hat{B} \right| \hat{\mathbbm{P}}^\dagger_1 \cdots \hat{\mathbbm{P}}^{\dagger}_N\right|0\right\rangle&=&\left\langle 0\left|\left[\hat{P}_1 \cdots \hat{P}_N, \hat{A}^\dagger \hat{B} \right]\hat{\mathbbm{P}}^{\dagger}_1 \cdots \hat{\mathbbm{P}}^{\dagger}_N\right|0\right\rangle\nonumber\\
&=& - \sum_{k=1}^N \frac{1}{2}{\rm tr}(p_{A} p_k)
\left\langle0 \left|\hat{P}_1 \cdots \hat{P}_{k-1} \hat{B} \hat{P}_{k+1}\cdots \hat{P}_N \left| \hat{\mathbbm{P}}^{\dagger}_1 \cdots \hat{\mathbbm{P}}^{\dagger}_N\right.\right |0\right\rangle\nonumber\\
&& + \sum_{k=2}^N \sum_{i=1}^k
\left\langle 0\left|\hat P_1 \cdots\hat{P}_{i-1} \hat{\mathcal P}_{i,k} \hat{P}_{i+1} \cdots \hat{P}_{k-1} \hat{B} P_{k+1} \cdots\hat{P}_N \left| \hat{\mathbbm{P}}^\dagger_1 \cdots \hat{\mathbbm{P}}^\dagger_N\right. \right|0\right\rangle.
\end{eqnarray}

Similarly, from Eq. (\ref{eq-con-ppqq}), one obtains matrix element of particle-hole interaction, $\hat{Q} \hat{Q}^\dagger$,
\begin{eqnarray}\label{ME-QQ}
\left\langle 0\left|\hat{P}_1 \cdots \right.\right.&\hat{P}_N &\left.\left.\left| \hat{Q} \hat{Q}^\dagger \right| \hat{\mathbbm{P}}^{\dagger}_1 \cdots \hat{ \mathbbm{P}}^{\dagger}_N\right|0\right\rangle=\left\langle 0\left|\left[\hat{P}_1 \cdots \hat{P}_N, \hat{Q} \hat{Q}^\dagger \right] \hat{\mathbbm{P}}^{\dagger}_1 \cdots \hat{ \mathbbm{P}}^{\dagger}_N\right|0\right\rangle\nonumber\\
&=& \sum_{k,i=1}^{N,k\neq i}
\left\langle 0\left|\hat P_1 \cdots
\right.\right.
\left\{
\begin{array}{cc}
\hat{P}_{i-1} \hat{\mathcal P}_i \hat{P}_{i+1} \cdots \hat{P}_{k-1} \hat{\mathcal P}_k \hat{P}_{k+1}&i<k\\
\hat{P}_{k-1} \hat{\mathcal P}_k \hat{P}_{k+1}\cdots \hat{P}_{i-1} \hat{\mathcal P}_i \hat{P}_{i+1} &k<i\\
\end{array}
\right\}
\left.\left.
\cdots \hat P_N \left|\hat{\mathbbm{P}}^{\dagger}_1 \cdots \hat{\mathbbm{P}}^{\dagger}_N\right.\right|0\right\rangle\nonumber\\
&& +\sum_{k=1}^{N}
\left\langle 0\left| \hat{P}_1 \cdots \hat{P}_{k-1} \hat{\mathfrak{P}}_k \hat{P}_{k+1} \cdots \hat{P}_N \left| \hat{\mathbbm{P}}^{\dagger}_1 \cdots \hat{\mathbbm{P}}^{\dagger}_N\right.\right|0\right\rangle,
\end{eqnarray}
where $\hat{\mathcal P}_i$, $\hat{\mathcal P}_k$, and $\hat{\mathfrak{P}}_k$ are given in Eq. (\ref{eq-ppqq-ope-def}).

\subsection{Odd-nucleon system}\label{sec-odd}

As discussed above, once the overlaps and matrix elements of one- and two-body operators for even-nucleon systems are obtained, results for odd-mass systems are readily obtained
in terms of those of even-nucleon systems. From Eq. (\ref{eqn:single-particle-operator-commutator}), one obtains overlap of an odd-mass nucleus in nucleon-pair basis states as below,
\begin{equation}\label{eq-ove-o}
\begin{aligned}
\left\langle0 \left|\hat{a} \hat{P}_1\cdots \hat{P}_N
\left| \hat{b}^\dagger \hat{\mathbbm{P}}^{\dagger}_1 \cdots \hat{\mathbbm{P}}^{\dagger}_N\right.\right|0\right\rangle=& (\vec{a} \cdot \vec{b}) \left\langle 0\left| \hat{P}_1 \cdots \hat{P}_N
\left| \hat{\mathbbm{P}}^{\dagger}_1 \cdots \hat{\mathbbm{P}}^{\dagger}_{N} \right.\right|0\right\rangle- \left\langle 0\left| \hat{P}_1 \cdots \hat{P}_N \left| \hat{\mathbb{Q}}\right | \hat{\mathbbm{P}}^{\dagger}_1 \cdots \hat{\mathbbm{P}}^{\dagger}_{N} \right|0\right\rangle,
\end{aligned}
\end{equation}
where $\hat{\mathbb{Q}}$ has structure matrix $ \mathbb{Q} = \vec{b} \vec{a}^\top$. On the right hand side of the above equation, the formula of the first term is given in Eq. (\ref{eq-ove-re}), and the formula of the second term is given in Eq. (\ref{eqn:ME-Q}). For the case of $N=0$,
\begin{eqnarray}
\left\langle \hat a\left|\hat b^\dagger\right.\right\rangle=\vec a^{\top}\vec b~. \nonumber
\end{eqnarray}

Similarly, we preset matrix elements of $\hat Q$, $\hat A^{\dagger} \hat{B}$, and $\hat Q\hat Q^{\dagger}$ for an odd-nucleon system, in terms of overlaps of Eq. (\ref{eq-ove-o}). By using Eq. (\ref{eq-con-ppq}), one obtains $\hat Q$ matrix element of odd-nucleon system as below,
\begin{eqnarray}\label{eqn:ME-Q-odd}
\left\langle 0\left|\hat{a} \hat{P}_1 \cdots \hat{P}_N \left| \hat{Q} \right| \hat{b}^{\dagger} \hat{\mathbbm{P}}^{\dagger}_1\right.\right.& \cdots& \left.\left.\hat{\mathbbm{P}}^{\dagger}_N \right|0\right\rangle=\left\langle 0\left|\left[\hat{a} \hat{P}_1 \cdots \hat{P}_N, \hat{Q} \right] \hat{b}^{\dagger} \hat{\mathbbm{P}}^{\dagger}_1 \cdots \hat{\mathbbm{P}}^{\dagger}_N \right|0\right\rangle\nonumber\\
&=&\left\langle 0\left| \hat d \hat{P}_1 \cdots \hat{P}_N \left| \hat{b}^\dagger \hat{\mathbbm{P}}^{\dagger}_1\cdots\mathbbm{P}^{\dagger}_{N} \right.\right|0\right\rangle+\sum_{k=1}^N\left\langle 0\left|\hat{a} \cdots \hat{P}_{k-1} \hat{\mathcal P}_k \hat{P}_{k+1} \cdots \hat{P}_N \left| \hat{b}^{\dagger} \cdots \hat{\mathbbm{P}}^{\dagger}_{N}\right.\right|0\right\rangle,
\end{eqnarray}
where $\hat d$ and $\hat{\mathcal P}_k$ is given in Eq. (\ref{eq-ppq-ope-def}). By using Eq. (\ref{eq-con-pppp}), one obtains the matrix elements of $\hat A^\dagger \hat{B} $ as follows.
\begin{eqnarray}\label{eqn:ME-AA-odd}
\left\langle 0\left|\hat{a} \hat{P}_1 \cdots \right.\right.&\hat{P}_N
&\left.\left.\left| \hat{A}^\dagger \hat{B}
\right| \hat{b}^\dagger \hat{\mathbbm{P}}^\dagger_1 \cdots \hat{\mathbbm{P}}^{\dagger}_N\right|0\right\rangle=\left\langle 0\left|\left[\hat{a} \hat{P}_1 \cdots \hat{P}_N ,\hat{A}^\dagger \hat{B}
\right] \hat{b}^\dagger \hat{\mathbbm{P}}^\dagger_1 \cdots \hat{\mathbbm{P}}^{\dagger}_N\right|0\right\rangle\nonumber\\
&= &- \sum_{k=1}^N \frac{1}{2}{\rm tr}(p_{A} p_k) \left\langle 0\left| \hat{a} \hat{P}_1 \cdots \hat{P}_{k-1} \hat{B} \hat{P}_{k+1} \cdots \hat{P}_N \left| \hat{b}^\dagger \hat{\mathbbm{P}}^{\dagger}_1
\hat{\mathbbm{P}}^{\dagger}_2 \cdots \hat{\mathbbm{P}}^{\dagger}_N \right.\right|0\right\rangle\nonumber\\
& &+ \sum_{k=2}^N \sum_{i=1}^k
\left\langle 0\left| \hat{a} \hat P_1\cdots \hat{P}_{i-1} \hat{\mathcal P}_{i,k}\hat{P}_{i+1} \cdots \hat{P}_{k-1} \hat{B} P_{k+1}\cdots
\hat{P}_N \left| \hat{b}^\dagger \hat{\mathbbm{P}}^\dagger_1 \cdots \hat{\mathbbm{P}}^\dagger_N \right.\right|0\right\rangle\nonumber\\
&& + \sum_{k=1}^N
\left\langle 0\left|\hat{e}_k\hat P_1 \cdots \hat{P}_{k-1} \hat{B} P_{k+1} \cdots \hat{P}_N\left | \hat{b}^\dagger \hat{\mathbbm{P}}^\dagger_1 \hat{\mathbbm{P}}^\dagger_2 \cdots \hat{\mathbbm{P}}^\dagger_N \right.\right|0\right\rangle,
\end{eqnarray}
where $\hat{\mathcal P}_{i,k}$ and $\vec{e}_k $ are given in Eq. (\ref{eq-ppp-ope-def}). By using Eq. (\ref{eq-con-ppqq}), one obtains matrix elements of $\hat Q\hat Q^\dagger$ as follows.
\begin{eqnarray}\label{eqn:ME-QQ-odd}
&&\left\langle 0\left| \hat{a} \hat{P}_1 \cdots \hat{P}_N \left| \hat{Q}\hat{Q}^\dagger
\right|\hat{b}^{\dagger} \hat{\mathbbm{P}}^{\dagger}_1 \cdots \hat{\mathbbm{P}}^{\dagger}_N\right| \right\rangle=\left\langle 0\left|\left[ \hat{a} \hat{P}_1 \cdots \hat{P}_N , \hat{Q}\hat{Q}^\dagger
\right]\hat{b}^{\dagger} \hat{\mathbbm{P}}^{\dagger}_1 \cdots \hat{\mathbbm{P}}^{\dagger}_N\right| \right\rangle\nonumber\\
&&=\left\langle 0\left| \hat g \hat{P}_1 \cdots \hat{P}_N \left| \hat{b}^\dagger \hat{\mathbbm{P}}^{\dagger}_1 \cdots \hat{\mathbbm{P}}^{\dagger}_{N}\right.\right|0\right\rangle+\sum_{k=1}^N\left\langle 0\left| \hat e \hat{P}_1 \cdots \hat{P}_{k-1} \mathcal P_k \hat{P}_{k+1}\cdots \hat{P}_N \left|\hat{b}^\dagger \hat{\mathbbm{P}}^{\dagger}_1 \cdots \hat{\mathbbm{P}}^{\dagger}_{N}\right.\right|0\right\rangle\nonumber\\
&&+\sum_{i=1}^N\left\langle 0\left|\hat f \hat{P}_1 \cdots \hat{P}_{k-1} \mathcal P_i \hat{P}_{k+1}\cdots \hat{P}_N \left| \hat{b}^\dagger \hat{\mathbbm{P}}^{\dagger}_1 \cdots \hat{\mathbbm{P}}^{\dagger}_{N}\right.\right|0\right\rangle+\sum_{k=1}^N\left\langle 0\left| \hat{a} \hat{P}_1 \cdots \hat{P}_{k-1} \hat{\mathfrak{P}}_k \hat{P}_{k+1} \cdots \hat{P}_N \left| \hat{b}^\dagger \hat{\mathbbm{P}}^{\dagger}_1 \cdots \hat{\mathbbm{P}}^{\dagger}_{N}\right.\right|0\right\rangle\nonumber\\
&&+\sum_{k,i=1}^{N,k\neq i} \left\langle 0\left|\hat{a}\hat P_1\cdots
\right.\right.
\left\{
\begin{array}{cc}
\hat{P}_{i-1} \mathcal P_i \hat P_{i+1} \cdots \hat{P}_{k-1} \mathcal P_k \hat{P}_{k+1}&i<k\\
\hat{P}_{k-1} \mathcal P_k \hat{P}_{k+1}\cdots\hat{P}_{i-1} \mathcal P_i\hat P_{i+1} &k<i\\
\end{array}
\right\}
\left.\left.
\cdots \hat P_N\left| \hat{b}^\dagger \cdots \hat{\mathbbm{P}}^{\dagger}_{N}\right.\right|0\right\rangle,
\end{eqnarray}
where $\mathcal P_k$, $\mathcal P_i$, $\hat{\mathfrak{P}}_k$, $\hat e$, $\hat f$, and $\hat g$ are given in Eq. (\ref{eq-ppqq-ope-def}),

\end{widetext}

\section{Nucleon-pair approximation with the matrix representation}\label{sec-mat-npa}

In this section, we restrict our nucleon pairs to those of the NPA defined in Refs. \cite{npa-for-chen-3,npa-for-zhao,npa-iso-fu,npa-ph-cheng}. In this case calculations by using formulas in the last section are equivalent to the NPA calculations. Therefore we call our formulation the matrix representation of the nucleon-pair approximation.

In the NPA, collective nucleon pair operator with given angular-momentum $r$ is
\begin{equation}\label{eq-jnpa-pair-definition}
\begin{aligned}
&A^{r\dagger}_m = \sum_{a b} y(abr)A^{r\dag}_m (ab),\\
&A^{r\dag}_m(ab) =\left( \hat c^{\dagger}_a \times \hat c^{\dagger}_b \right)^{r}_m, \\
&A^{r}_m = \sum_{a b} y(abr) A^{r}_m (ab),\\
&A^{r}_m(ab) =\left(A^{r\dag}_m(ab)\right)^{\dagger}=-\left(\hat c_a \times \hat c_b \right)^{r}_m,
\end{aligned}
\end{equation}
where $r$ and $m$ are the angular momentum of the pair and its projection to principal axis, respectively. $a$ and $b$ represent the quantum numbers of spherical single-particle basis, \{$nlj$\}, the symbol $\times$ corresponds to angular-momentum couplings, and $y(abr)$ is the structural coefficient of the collective pair. $y(abr)$ has anti-symmetric property as $y(abr)=-(-)^{r-j_a-j_b}y(bar)$. In the form of our collective pair in Eq. (\ref{eq-pair-def}), one has
\begin{equation}
\begin{aligned}
A^{r\dagger} \rightarrow \hat P^{\dagger} = \sum_{am_a,bm_b}p(am_a,bm_b)\hat c^{\dagger}_{am_a}c^{\dagger}_{bm_b}
\end{aligned}
\end{equation}
with
\begin{equation}\label{eq-p2y}
p(am_a,bm_b)=2y(abr)\langle j_am_aj_bm_b|rm\rangle,
\end{equation}
where $\alpha\equiv am_a$ and $\beta\equiv bm_b$, $m_a$ and $m_b$ are the projections of $j_a$ and $j_b$ to principal axis, and $\langle j_am_aj_bm_b|rm\rangle$ is the Clebsch-Gorden coefficients. One easily sees that $p(am_a,bm_b) = - p(bm_b, am_a)$, and that the operator
\begin{eqnarray}
\hat P^{\dagger} \equiv A^{r\dagger}~. \nonumber
\end{eqnarray}

In the NPA with isospin, the collective nucleon pair is
\begin{equation}
\begin{aligned}
&A^{r\mathbbm t\dagger}_{m\tau} = \sum_{a b} y(abr\mathbbm t)A^{r\mathbbm t\dag}_{m\tau} (ab),\\
&A^{r\mathbbm t\dag}_{m\tau}(ab) =\left( \hat c^{\dagger}_a \times \hat c^{\dagger}_b \right)^{r\mathbbm t}_{m\tau}, \\
&A^{r\mathbbm t}_{m\tau} = \sum_{a b} y(abr\mathbbm t) A^{r\mathbbm t}_{m\tau} (ab),\\
&A^{r\mathbbm t}_{m\tau}(ab) =\left(A^{r\mathbbm t\dag}_{m\tau}(ab)\right)^{\dagger}=-\left(\hat c_a \times \hat c_b \right)^{r\mathbbm t}_{m\tau},
\end{aligned}
\end{equation}
where $\mathbbm t$ and $\tau$ is the isospin and its projection to the principal axis, and structural coefficients $y(abr\mathbbm t)$ has anti-symmetry of $y(bar\mathbbm t)=(-)^{r-j_a-j_b+\mathbbm t}y(abr\mathbb)$. Similarly, we define our collective pair
\begin{equation}
\begin{aligned}
\hat P^{\dagger}=\sum_{am_a\tau_a;bm_b\tau_b}p(am_a\tau_a,am_b\tau_b)\hat c^{\dagger}_{am_a\tau_a}\hat c^{\dagger}_{bm_b\tau_b}
\end{aligned}
\end{equation}
with
\begin{eqnarray}
p(am_a\tau_a,bm_b\tau_b)&=&2y(abr\mathbbm t)\left\langle j_am_aj_bm_b|rm\right\rangle \nonumber\\
&&\times\left\langle\frac{1}{2}\tau_a\frac{1}{2}\tau_b|\mathbbm t\tau\right\rangle,
\end{eqnarray}
where $\alpha\equiv am_a\tau_a$ and $\beta\equiv bm_b\tau_b$, and $\tau_a$ and $\tau_b$ are the single-particle isospin projections. The structural coefficients $p(am_a\tau_a,bm_b\tau_b)$ satisfies the requirement
\begin{eqnarray}
p(bm_b\tau_b,am_a\tau_a) = - p(am_a\tau_a, b m_b\tau_b) ~, \nonumber
\end{eqnarray}
and the collective nucleon-pair operator such defined is equivalent to $A^{r\mathbbm t\dagger}$.

The one-body operator in the NPA is in following form,
\begin{equation}
\begin{aligned}
Q^k_{\kappa}&=\sum_{ab}{q}(abk)(c^{\dagger}_a\times \tilde c_b)^k_{\kappa},\\
Q^{k\mathbbm t}_{\kappa\tau}&=\sum_{ab}{q}(abk\mathbbm t)(c^{\dagger}_a\times \tilde c_b)^{k\mathbbm t}_{\kappa\tau},
\end{aligned}
\end{equation}
where $ q(abk)$ and $ q(abk\mathbbm t)$ correspond to structural coefficients of one-body operator in the NPA with and without isospin, respectively. $\tilde c_{b}$ is the time-reversal operator of $c_{b}$ with $\tilde c_{j_bm_b}=(-)^{j_b-m_b}c_{j_b-m_b}$ and $\tilde c_{j_bm_b\tau_b}=(-)^{j_b-m_b+1/2-\tau_b}c_{j_b-m_b-\tau_b}$. The one-body operator is readily rewritten in the form of Eq. (\ref{eq-q}) as below,
\begin{equation}\label{eq-q2q}
\begin{aligned}
Q^k_{\kappa}&=\sum_{am_a,bm_b}q(am_a,bm_b)\hat c^{\dagger}_{am_a}\hat c_{bm_b},\\
Q^{k\mathbbm t}_{\kappa\tau}&=\sum_{am_a\tau_a,bm_b\tau_b}q(am_a\tau_a,bm_b\tau_b)\hat c^{\dagger}_{am_a\tau_a}\hat c_{bm_b\tau_b},\\
\end{aligned}
\end{equation}
with
\begin{eqnarray}
&& q(am_a,bm_b) = (-)^{j_b+m_b} q(abk) \left\langle j_am_aj_b-m_b|k\kappa\right\rangle~, \nonumber \\
&& q(am_a\tau_a,bm_b\tau_b) = (-)^{j_b+m_b+\frac{1}{2}+\tau_b} q(abk\mathbbm t) \nonumber \\
&& ~~~~~~~~~~~~~~~~~~ \left\langle j_am_aj_b-m_b|k\kappa\right\rangle \left\langle \frac{1}{2}\tau_a\frac{1}{2}-\tau_b|\mathbbm t\tau\right\rangle~. \nonumber
\end{eqnarray}
In most cases, one-body operators in the NPA are hermitian or anti-hermitian, corresponding to
\begin{equation}
\begin{aligned}
\left(Q^k_{\kappa}\right)^{\dagger}=&\pm (-)^{k-\kappa}Q^k_{-\kappa},\\
\left(Q^{k\mathbbm t}_{\kappa\tau}\right)^{\dagger}=&\pm (-)^{k-\kappa+\mathbbm t -\tau}Q^{k\mathbbm t}_{-\kappa-\tau},\\
\end{aligned}
\end{equation}
and
\begin{equation}
\begin{aligned}
q(am_a,bm_b)=&\pm q(bm_b,am_a),\\
q(am_a\tau_a,bm_b\tau_b)=&\pm q(bm_b\tau_b,am_a\tau_a),
\end{aligned}
\end{equation}
respectively.

By using notations of Eqs. (\ref{eq-p2y}) and (\ref{eq-q2q}), the shell-model Hamiltonian of like nucleons are
\begin{equation}\label{eq-npa-h-pp-nn}
\begin{aligned}
H_{\rm like}&=\sum_a\varepsilon_a \sum_{m_a\tau_a}c^{\dagger}_{j_am_a(\tau_a)}c_{j_am_a(\tau_a)}\\
&+\sum_{L({\mathbbm t}),i}G^i_{L({\mathbbm t})}\sum_{m(\tau)}A^{L({\mathbbm t})\dagger}_{m(\tau)}(y_i)A^{L({\mathbbm t})}_{m(\tau)}(y_i)\\
&+\sum_{ k({\mathbbm t}),i}F^i_{k({\mathbbm t})}\sum_{\kappa\tau}Q^{k({\mathbbm t})}_{\kappa(\tau)}( q_i)Q^{k({\mathbbm t})\dagger}_{\kappa(\tau)}(q_i),
\end{aligned}
\end{equation}
where the parentheses of $\mathbbm t$ and $\tau$ means that the isospin degree of freedom can be suppressed for interactions between like valence nucleons. The three terms in this Hamiltonian corresponds to one-body term, multipole pairing interaction, and multipole-multipole interaction, respectively, where $\varepsilon_{a}$, $G$, and $F$ are the single-particle energy and interaction strengths of multipole pairing and multipole-multipole interaction. We distinguish $A^{L({\mathbbm t})\dagger}_{m(\tau)}A^{L({\mathbbm t})}_{m(\tau)}$ (and $Q^{k({\mathbbm t})}_{\kappa(\tau)}Q^{k({\mathbbm t})\dagger}_{\kappa(\tau)}$ terms) with different $y_i$ (or $q_i$) matrices but the same $L({\mathbbm t})$ (or $k({\mathbbm t})$) quantum number(s).
The one-body single-particle term can be rewritten in the form of Eq. (\ref{eq-q}),
\begin{eqnarray}
&& \sum_a\varepsilon_a \sum_{m_a\tau_a}c^{\dagger}_{j_am_a(\tau_a)}c_{j_am_a(\tau_a)} \nonumber \\
& = & \sum_{\alpha \beta} q(\alpha \beta) \hat{c}^{\dagger}_{\alpha} \hat{c}_{\beta} ~, \nonumber
\end{eqnarray}
where
\begin{eqnarray}
q(\alpha \beta) \equiv q\left(am_a(\tau_a),bm_b(\tau_b)\right)=\delta_{ab}\delta_{m_am_b}(\delta_{\tau_a\tau_b})\varepsilon_{a} ~. \nonumber
\end{eqnarray}

The proton-neutron interaction in the particle-hole channel is written as below,
\begin{equation}\label{eq-npa-h-pn}
H_{\pi\nu}=\sum_{k,i}F^i_{k}\sum_{\kappa}Q^{k}_{\kappa}(q_{\pi,i})(-)^{\kappa}Q^{k}_{-\kappa}(q_{\nu,i}),
\end{equation}
where the $\pi$ and $\nu$ correspond to proton and neutron degrees of freedom, respectively, and the label $i$ is introduced to distinguish $Q^{k}_{\kappa}(q_{\pi})(-)^{\kappa}Q^{k}_{-\kappa}(q_{\nu})$ interactions with the same $k$ but different $q_{\pi,i}$ and $q_{\nu,i}$ matrices of structural coefficients.

The NPA basis states are constructed in the following form,
\begin{equation}\label{eq-npa-bas}
\left.\left.A^{\dagger}_0A^{r_1(\mathbbm t_1)\dagger}_{m_1(\tau_1)}A^{r_2(\mathbbm t_2)\dagger}_{m_2(\tau_2)}\cdots A^{r_N(\mathbbm t_N)\dagger}_{m_N(\tau_N)}\right|0\right\rangle\\
\end{equation}
with
\begin{equation*}
\hat A^{\dagger}_0=\left\{
\begin{array}{cl}
1&{\rm ~for~an~even-even~nucleus}\\
\hat c^{\dagger}_{jm(\tau)} &{\rm ~for~an~odd~mass~ nucleus}
\end{array}
\right.,
\end{equation*}
where the parentheses of $\mathbbm t$ and $\tau$ means, again, that the isospin degree of freedom might be suppressed. The above basis states do not have given angular momentum, but diagonalization of the shell model Hamiltonian yields the same wave functions as the NPA calculations once the configuration space in the form of the above basis states is complete. The dimension of the above basis states equals to that of the NPA configuration space, as pointed out in Ref. \cite{mnpa-luo}.

From the discussion in this section, we have shown that the collective pair in the NPA, and one-body and two-body interactions of the nuclear shell model, can be formulated with the matrix representation. The formulas in Sec. \ref{sec-for-2} is applicable to the NPA calculation, with the basis states written in the form of Eq. (\ref{eq-npa-bas}). We see clearly that, the matrix representation of the NPA is formulated in a unified way, for both even-even nuclei and odd-mass nuclei, and both with and without isospin symmetry.

\section{Advantage of the NPA with matrix representation }\label{sec-alg}

In this section, we discuss the computational advantage of the matrix-represented NPA, from two aspects: overlap computation and recursion number.

\subsection{Overlap computation}\label{sec-per}
\begin{figure}
\includegraphics[angle=0,width=0.48\textwidth]{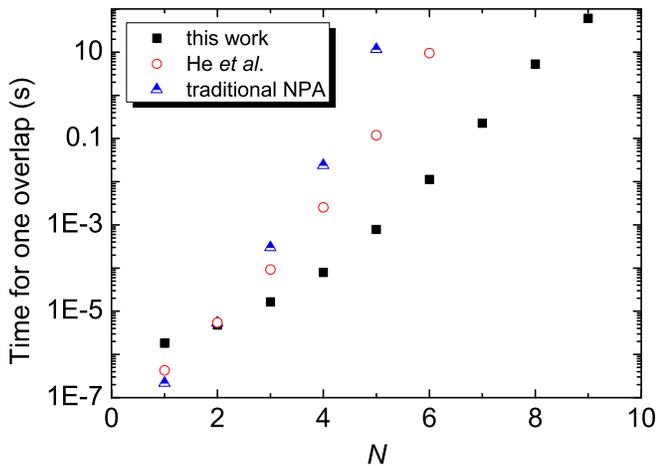}
\caption{(color online) Average computational time of one overlap versus nucleon-pair number $N$, for three approaches of the NPA,
exemplified by using $SD$-pair space in the $pf$ shell. ``traditional NPA'' corresponds to the NPA of Refs. \cite{npa-for-chen-3,npa-for-zhao},
``He $et~al.$" corresponds to the NPA of Ref. \cite{mnpa-luo}, and ``this work" corresponds to the NPA with matrix representation.
The calculations are carried out on a PC platform with CPU frequency 4.9GHz. }\label{fig-ove-tim-en}
\end{figure}

The computation of overlaps between basis states is the key part of all approaches of the NPA. In this subsection, we discuss the computational advantage of our matrix-represented NPA in calculations of overlaps between the basis states, in comparison with those in the NPA of Refs. \cite{npa-for-chen-3,npa-for-zhao} and of Ref. \cite{mnpa-luo}. This comparison is exemplified by using $SD$-pair approximation of the $pf$ shell.

In Fig. \ref{fig-ove-tim-en}, we plot the evolution of the average computational time of one overlap (in Seconds, denoted by ${\Gamma}$) with nucleon-pair number $N$.
According to Fig. \ref{fig-ove-tim-en}, the value of $\Gamma$ for the NPA of Refs. \cite{npa-for-chen-3,npa-for-zhao} is the smallest when $N=1$;
and the value of $\Gamma$ is almost the same for three approaches, Refs. \cite{npa-for-chen-3,npa-for-zhao}, Ref. \cite{mnpa-luo}, and the present NPA with matrix representation when $N=2$. For $N \ge 3$, the value of $\Gamma$ for the NPA with matrix representation is substantially superior to that for the NPA of Refs. \cite{npa-for-chen-3,npa-for-zhao} and of Ref. \cite{mnpa-luo}. For instance, for $N=5$, the value of $\Gamma$ by the using approaches of Refs. \cite{npa-for-chen-3,npa-for-zhao} is larger by about 2 orders than that of Ref. \cite{mnpa-luo}, and by about 4 orders than that of the NPA with matrix representation suggested in this work.

It is worthy to understand the reason why the value of $\Gamma$ exhibits the above behavior. In the case of $N=1$, the NPA of Refs. \cite{npa-for-chen-3,npa-for-zhao} calculates the overlaps by a very compact formula,
\begin{eqnarray}
\langle r_1, J_1 | s_1, J_1 \rangle = 2 \delta_{r_1 s_1} \delta_{r_1 J_1} \delta_{s_1 J_1} \sum_{ab} y(abr_1)y(abs_1) ~; \nonumber
\end{eqnarray}
for $N=2$, the overlaps are given as four-dimensional summation of nine-$j$ coefficients, as shown in Eq. (A2) of Ref. \cite{npa-for-chen-3}; for $N \ge 3$, the overlaps are given in terms of overlaps for $N-1$ nucleon pairs, with new intermediate angular momenta and new pair structural coefficients due to angular-momentum recouplings and commutations. Therefore calculation of overlaps become more and more difficult when $N$ is larger than 5, by using the NPA of Refs. \cite{npa-for-chen-3,npa-for-zhao}.

The approach of Ref. \cite{mnpa-luo} avoids the heavy calculation of summation over new intermediate angular momenta $J_i$, yet suffers
the couplings and recouplings to calculate the structural coefficients arising from the commutations. In the NPA of Ref. \cite{mnpa-luo}, as in Refs. \cite{npa-for-chen-3,npa-for-zhao}, computation of commutators between collective pair and one-body operator involves of three-folded summation over six-$j$ symbols [for example, see Eqs. (2.10a) and (4.5a) of Ref. \cite{npa-for-zhao}, or Eqs. (16) and (17) of Ref. \cite{mnpa-luo}].

In the NPA with matrix representation, as shown in Eqs. (\ref{eq-ove-re}), (\ref{eq-ppp-pre}), and (\ref{eq-ove-o}), the overlaps of $N$ nucleon pair are written in terms of those of $N-1$ nucleon pair, with only products of matrices involved; couplings and recouplings of angular momenta, the challenges of summation over new intermediate angular momenta $J_i$ and (numerous and iterative) calculations of new pair structural coefficients are avoided. Furthermore, modern computing architecture favors the matrix-product operation, which is well optimized and automatically paralleled in most cases (see matrix operations in Refs. \cite{blas, mkl}). This is one of the advantages of the present approach.

\subsection{Recursive number}\label{sec-tim}

All NPA calculations are performed recursively. Therefore smaller numbers of recursion are favorable in the NPA calculations. Below we enumerate the number
of recursions involved in the evaluation of overlap for $N$ nucleon-pair basis states in terms of overlaps of $(N-1)$ nucleon-pair basis states.

\begin{figure}
\includegraphics[angle=0,width=0.48\textwidth]{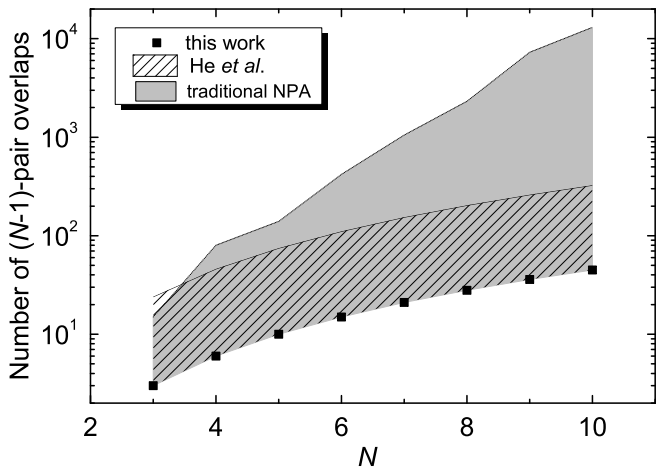}
\caption{ Number of overlaps with $(N-1)$ nucleon pairs for an overlap with $N$ nucleon pairs, versus nucleon-pair number $N$, for the NPA calculations of Refs. \cite{npa-for-chen-3,npa-for-zhao} (shadow in grey, labeled by ``traditional NPA"), of Ref. \cite{mnpa-luo} (stripes, labeled by
``He {\it et al.}"), and of the present formulation (solid squares in black, labeled by ``this work"). See the text for details. }\label{fig-re-cou}
\end{figure}

Let us first look at the approach of Refs. \cite{npa-for-chen-3,npa-for-zhao}. In this approach, new intermediate quantum numbers arise from angular-momentum recouplings of basis states and nucleon-pairs, i.e., $L$ quantum numbers therein \cite{npa-for-chen-3,npa-for-zhao}. In this case, it is difficult to enumerate the number of overlaps with $(N-1)$ nucleon pairs analytically; here we perform this enumeration by using our computer code, and plot the numbers versus pair number $N$ in Fig. \ref{fig-re-cou}, denoted by using shadows in grey, for an $SD$-pair configuration in the $pf$ shell. According to our enumeration, as plotted in Fig. \ref{fig-re-cou} for overlap with $N =6$, for example, one has to calculate up to a few hundreds of overlaps of basis states with $N=5$.

The NPA approach of Ref. \cite{mnpa-luo} involves of smaller recursion numbers of overlaps with $N-1$ nucleon pairs than the approach of Refs. \cite{npa-for-chen-3,npa-for-zhao} in the overlaps of $N$ nucleon pairs. In the overlap $\left\langle 0\left| D_{1}D_2\cdots D_i \cdots D_{N}\left|D^\dagger_{1}D^\dagger_2\cdots D^{\dagger}_j \cdots D^\dagger_{N} \right.\right|0\right\rangle$ of Ref. \cite{mnpa-luo},
each $\{i,k\}$ combination includes 7 overlaps of $(N-1)$-pair system, with 7 possible intermediate pairs of $A^{r^{\prime}_i=0\sim 6}$. Totally there are $N(7N-5)/2$ overlaps with $(N-1)$ nucleon pairs are involved in calculations of overlaps with $N$ nucleon pairs. The numbers of overlaps with $(N-1)$ nucleon pairs for various nucleon-pair basis states
are plotted by using stripes in Fig. \ref{fig-re-cou}, with their maxima $N(7N-5)/2$.

In the NPA with matrix representation in this paper, all nucleon pairs including those given by double commutation, as shown in Eq. (\ref{eq-ppp-ope-def}), are represented by matrices, and there is no multiplicity of angular momentum arising from commutation. There are $N(N-1)/2$ overlaps of $(N-1)$ nucleon pairs with a new pair of $\hat {\mathcal P}_{i,k}$ given in Eq. (\ref{eq-ppp-ope-def}), and $N$ overlaps of $(N-1)$ nucleon pairs without new pairs. Totally one $N$-pair overlap involves $N(N+1)/2$ overlaps of $(N-1)$ nucleon pairs, as denoted by solid squares in black. This number is in general smaller than that in the NPA approach of Ref. \cite{mnpa-luo}.

In Fig. \ref{fig-re-cou} we plot the numbers of overlaps with $(N-1)$ nucleon pairs, versus nucleon-pair number $N$, for the NPA of Refs. \cite{npa-for-chen-3,npa-for-zhao}, the NPA of Ref. \cite{mnpa-luo} and the NPA with matrix representation suggested in this paper. One sees that the number of overlaps with $(N-1)$ nucleon pairs involved in an overlap of $N$ nucleon pairs for the matrix represented NPA is the smallest among all. Clearly, in cases of nucleon pairs with larger spins, the advantage of the matrix represented NPA is even more striking.

\section{summary}\label{sec-sum}

To summarize, in this paper we revisit the NPA formulation with the matrix representation. The nucleon-pair structural coefficients in this formulation are represented in terms of anti-symmetrized matrices. The overlaps, matrix elements of one-body operators and shell model Hamiltonian are calculated recursively as previous approaches \cite{npa-for-chen-3,npa-for-zhao,npa-iso-fu,npa-ph-cheng,mnpa-luo}.

The improvement in the present approach is attributed to the removal of angular-momentum couplings and recouplings throughout the computation. Without the angular-momentum couplings and re-couplings, the commutations of nucleon pair operators with one-body and two-body operators are now presented in terms of matrix products which well fit the architecture 
of modern computers, e.g., see Refs. \cite{blas,mkl}. Moreover, this improvement reduces the number of recursions substantially. Therefore the NPA computational efficiency is enhanced drastically. With a conventional laptop computer, one readily diagonalizes the shell model Hamiltonian of six valence proton pairs and six valence neutrons in the $SD$-pair truncated shell model space. Exemplified by $^{154}$Sm nucleus, it takes about 36 hours to calculate the low-lying states for the $SD$-pair approximation, by using a computer of 
CPU frequency 4.9GHz. Therefore with the matrix representation, studies of the rotational motion of heavy nuclei are now realizable within the framework of the nucleon-pair approximation of the nuclear shell model.

Another advantage of the present formulation is flexibility. The NPA calculations with and without isospin symmetry, and for odd and even nucleons,
are formulated on the same footing.

\noindent{\bf Conflict of interest}

The authors declare that they have no conflict of interest.

\acknowledgements

We thank Bing-Cheng He for discussions and communications, and Calvin W Johnson for constructive suggestions. Y. Lei is grateful for the financial support of the Sichuan Science and Technology Program (Grant No. 2019JDRC0017), and the Doctoral Program of Southwest University of Science and Technology (Grant No. 18zx7147). Y. Lu acknowledges support from the National Natural Science Foundation of China (Grant No. 11705100), the Youth Innovations and Talents Project of Shandong Provincial Colleges, and Universities (Grant No. 201909118), Higher Educational Youth Innovation Science and Technology Program Shandong Province (Grant No. 2020KJJ004). Y. M. Zhao thanks the National Natural Science Foundation of China (Grants No. 11975151, No. 11675101, and No. 11961141003) and MOE Key Lab for Particle Physics, Astrophysics and Cosmology for financial support.

\end{CJK*}
\bibliographystyle{apsrev4-2}
\bibliography{myref}

\end{document}